\documentclass{PoS}

\usepackage{cite}

\newcommand{\be}{\begin{equation}}
\newcommand{\ee}{\end{equation}}
\newcommand{\bea}{\begin{eqnarray}}
\newcommand{\eea}{\end{eqnarray}}

\newcommand{\bmp}[1]{\begin{minipage}{#1cm}}
\newcommand{\emp}{\end{minipage}}

\newcommand{\xv}{{\mathbf x}}
\newcommand{\half}{\frac{1}{2}}

\newcommand{\om}{\omega}

\newcommand{\bean}{\begin{eqnarray*}}
\newcommand{\eean}{\end{eqnarray*}}

\newcommand{\vecx}{{\mathbf x}}

\title{News from bottomonium spectral functions in thermal QCD}

\ShortTitle{Bottomonium spectral functions in thermal QCD}

\author{\speaker{Sam Offler}, Gert Aarts, Chris Allton, Jonas Glesaaen  \\
    Department of Physics, Swansea University, Swansea SA2 8PP, United Kingdom\\
     E-mail: 
     \email{\{s.p.offler.967106, g.aarts, c.allton\}@swansea.ac.uk, jonas@glesaaen.com}}

\author{Benjamin J\"ager\\
        CP3-Origins \& Danish IAS, Department of Mathematics and
	Computer Science, University of Southern Denmark, 5230 Odense M, Denmark \\
	E-mail: 
	\email{jaeger@cp3.sdu.dk}}

\author{Seyong Kim\\
        Department of Physics, Sejong University, Seoul 143-747, Korea\\
        E-mail: 
        \email{skim@sejong.ac.kr}}

\author{Maria Paola Lombardo\\
        INFN, Sezione di Firenze, 50019 Sesto Fiorentino (FI), Italy\\
        E-mail: 
        \email{Mariapaola.Lombardo@lnf.infn.it}}

\author{Sinead M.\ Ryan\\
	School of Mathematics, Trinity College, Dublin 2, Ireland\\
	E-mail: 
	\email{ryan@maths.tcd.ie}}

\author{Jon-Ivar Skullerud\\
        Dept.\ of Theoretical Physics, National University of Ireland Maynooth, County Kildare, Ireland\\
        E-mail: 
        \email{jonivar@thphys.nuim.ie}}

\abstract{
New results on bottomonium at nonzero temperature are presented, using the FASTSUM Generation 2L ensembles. Preliminary results for spectral function reconstruction using Kernel Ridge Regression, a machine learning technique,  are shown as well and compared to results from the Maximum Entropy Method.}

\FullConference{37th International Symposium on Lattice Field Theory - Lattice2019\\
		16-22 June 2019\\
		Wuhan, China}

\begin{document}

\section{Introduction}

Quarkonia, bound states of a heavy quark and anti-quark, are among the most-studied probes of the quark-gluon plasma, both experimentally and theoretically, see e.g.\ the reviews \cite{Aarts:2016hap,Rothkopf:2019ipj}.
In this contribution, we report on the latest preliminary results of the FASTSUM collaboration \cite{fastsum} for bottomonium, using lattice NRQCD on our new {\em Generation} 2L thermal ensembles.
These calculations, with $N_f=2+1$ flavours of Wilson-clover quark,  are performed on anisotropic lattices, with $\xi = a_s/a_\tau =3.453(6)$ and a fixed cutoff of $a_\tau^{-1}= 5.997(34)$ GeV, a pion mass of $m_\pi = 236(2)$ MeV, and a physical strange quark. Some details on the ensembles are given in Table \ref{tab2L} and a full discussion of Gen 2L is given in a companion contribution \cite{Aarts:2019hrg}. In particular, an estimate of the pseudocritical temperature from the renormalised chiral condensate is given by $T_{\rm pc}=162(1)$ MeV, which we will use as an indication for the crossover temperature.

The current study extends our previous Generation 1  \cite{Aarts:2010ek,Aarts:2011sm,Aarts:2012ka,Aarts:2013kaa}
 and especially Generation 2 \cite{Aarts:2014cda} analysis, which differs from the Gen 2L ensembles by having a heavier pion, $m_\pi = 384(4)$ MeV. Hence one of the objectives of this study is to investigate the role of the light quarks. A second objective is to explore new methods for extracting spectral information from Euclidean correlators and we will present some very preliminary results obtained using a machine learning approach, namely Kernel Ridge Regression.

\begin{table}[b]
\begin{center}
\begin{tabular}{|c|| ccccccc |}
\hline
$N_\tau$        & 256$^*$& 128  & 64    & 56    & 48    & 40    & 36 \\ \hline
$T$ [MeV]  	& 23    & 47    & 94    & 107   & 125   & 150   & 167 \\
$N_{\rm cfg}$ 	& 750   & 306   & 1041  & 1042  & 1123  & 1102  & 1119 \\
\hline
$N_\tau$        & 32    & 28    & 24    & 20    & 16    & 12    & 8 \\    \hline
$T$ [MeV]  		& 187   & 214   & 250   & 300   & 375   & 500   & 750\\
$N_{\rm cfg}$ 	& 1090  & 1031  & 1016  & 1030  & 1102  & 1267  & 1048  \\
\hline
\end{tabular}
\caption{Generation 2L ensembles, $m_\pi=236(2)$ MeV, lattice size $32^3 \times N_\tau$, spatial lattice spacing $a_s=0.1136(6)$ fm, temporal lattice spacing $a_\tau^{-1}=5.997(34)$ GeV, anisotropy $a_s/a_\tau=3.453(6)$. The choice of parameters and the ensemble at the lowest temperature are courtesy of the HadSpec collaboration \cite{Wilson:2015dqa}.
\label{tab2L}}
\end{center}
\end{table}

\section{NRQCD correlation functions}

\begin{figure}
\begin{center}
	\includegraphics[width=0.49\linewidth]{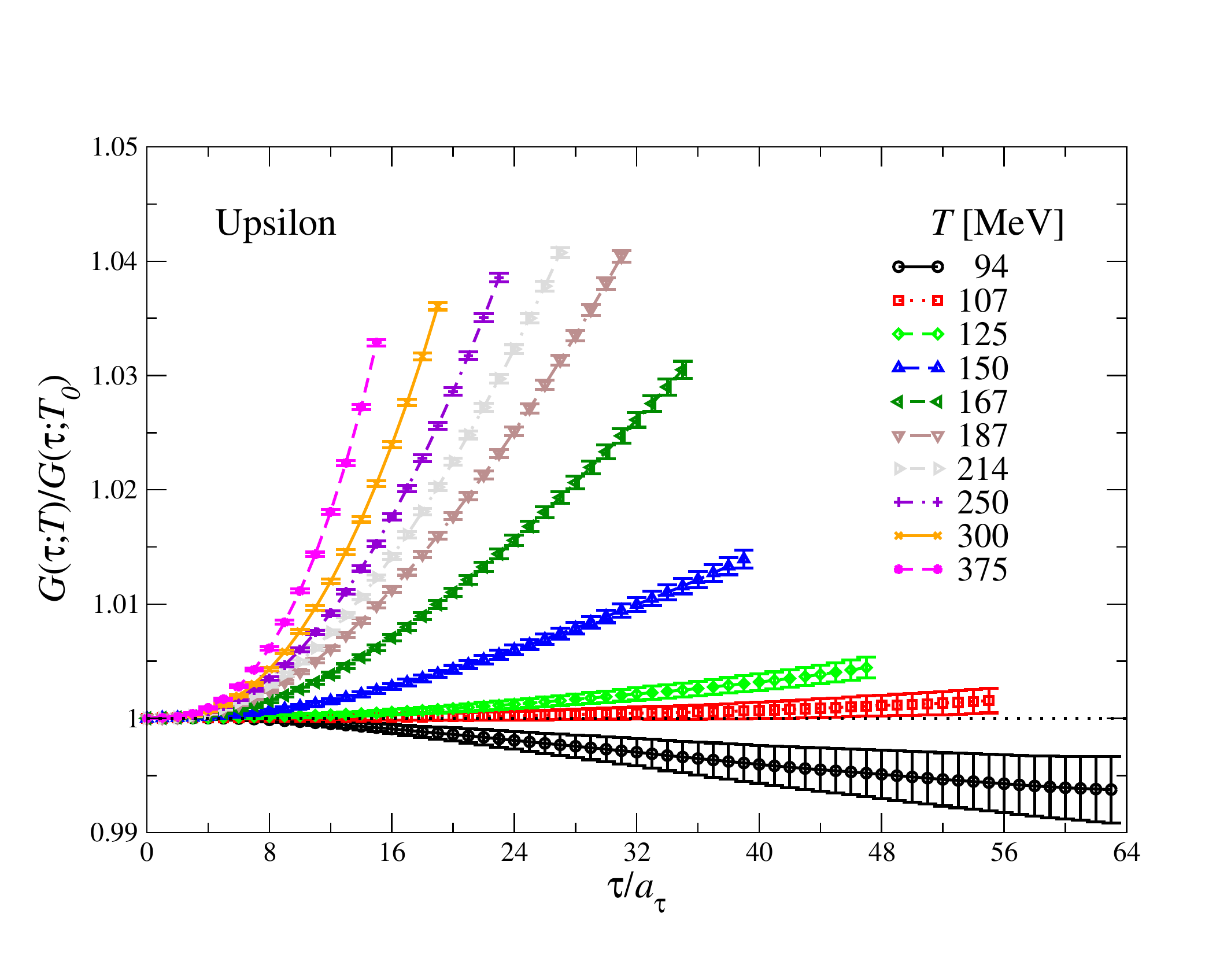}
	\includegraphics[width=0.49\linewidth]{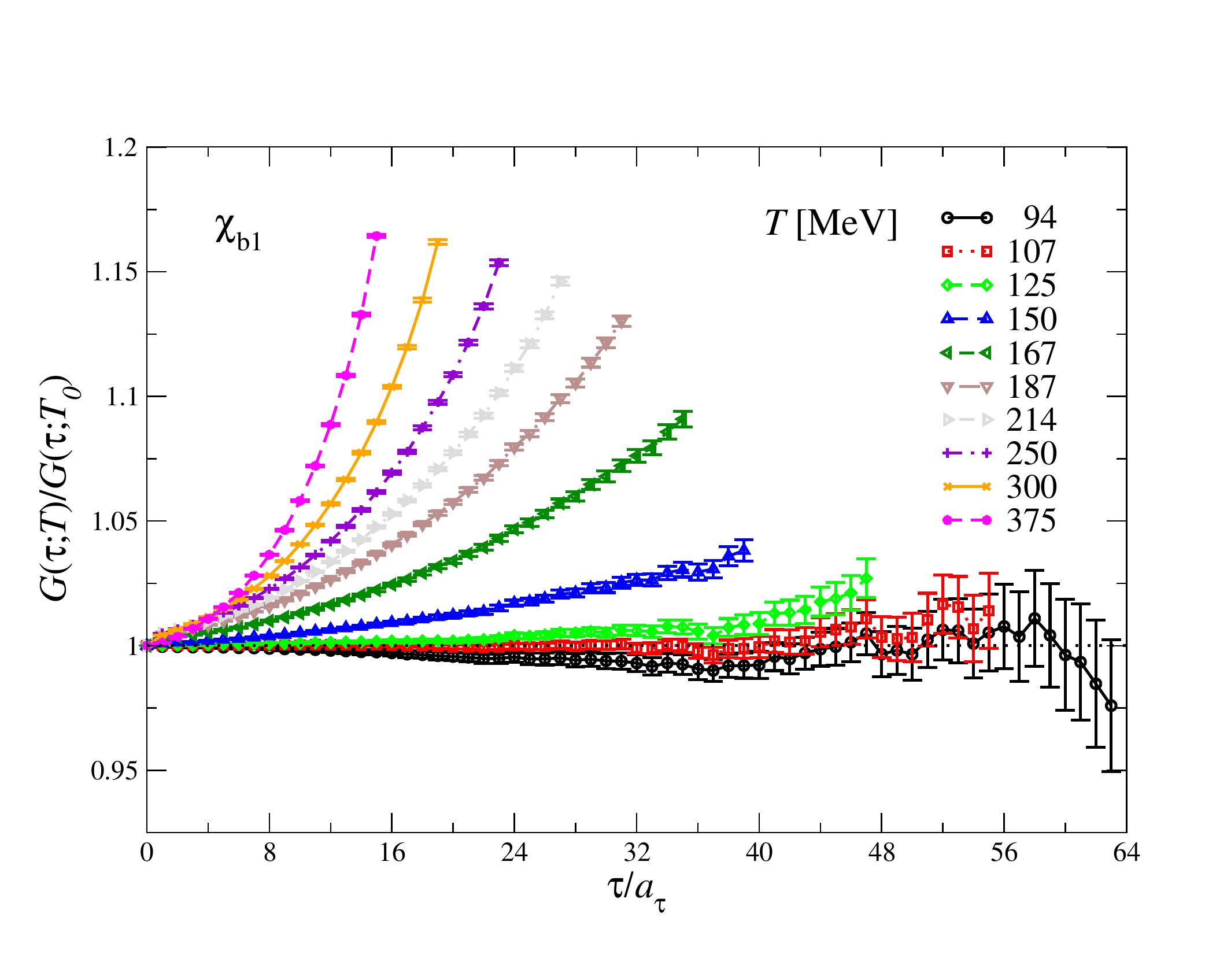}
	\caption{Thermal modification, $G(\tau; T)/G(\tau; T_0)$ with $T_0=47$ MeV, of the correlation functions in the $\Upsilon$ (left) and $\chi_{b1}$ (right) channels. Note the different vertical scale.
	\label{fig:ratio}}
\end{center}

\vspace*{-1cm}

\begin{center}
	\includegraphics[width=0.49\linewidth]{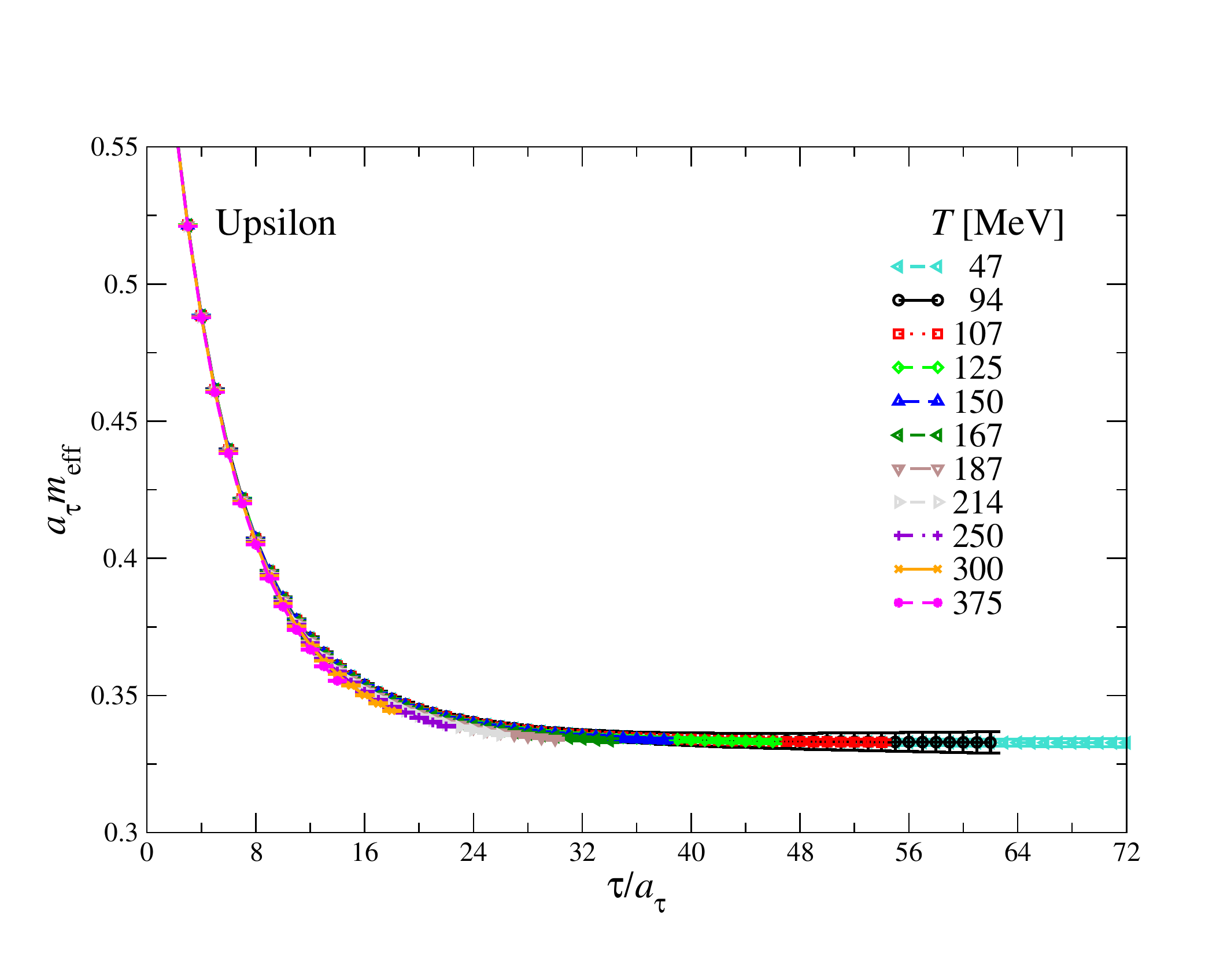}
	\includegraphics[width=0.49\linewidth]{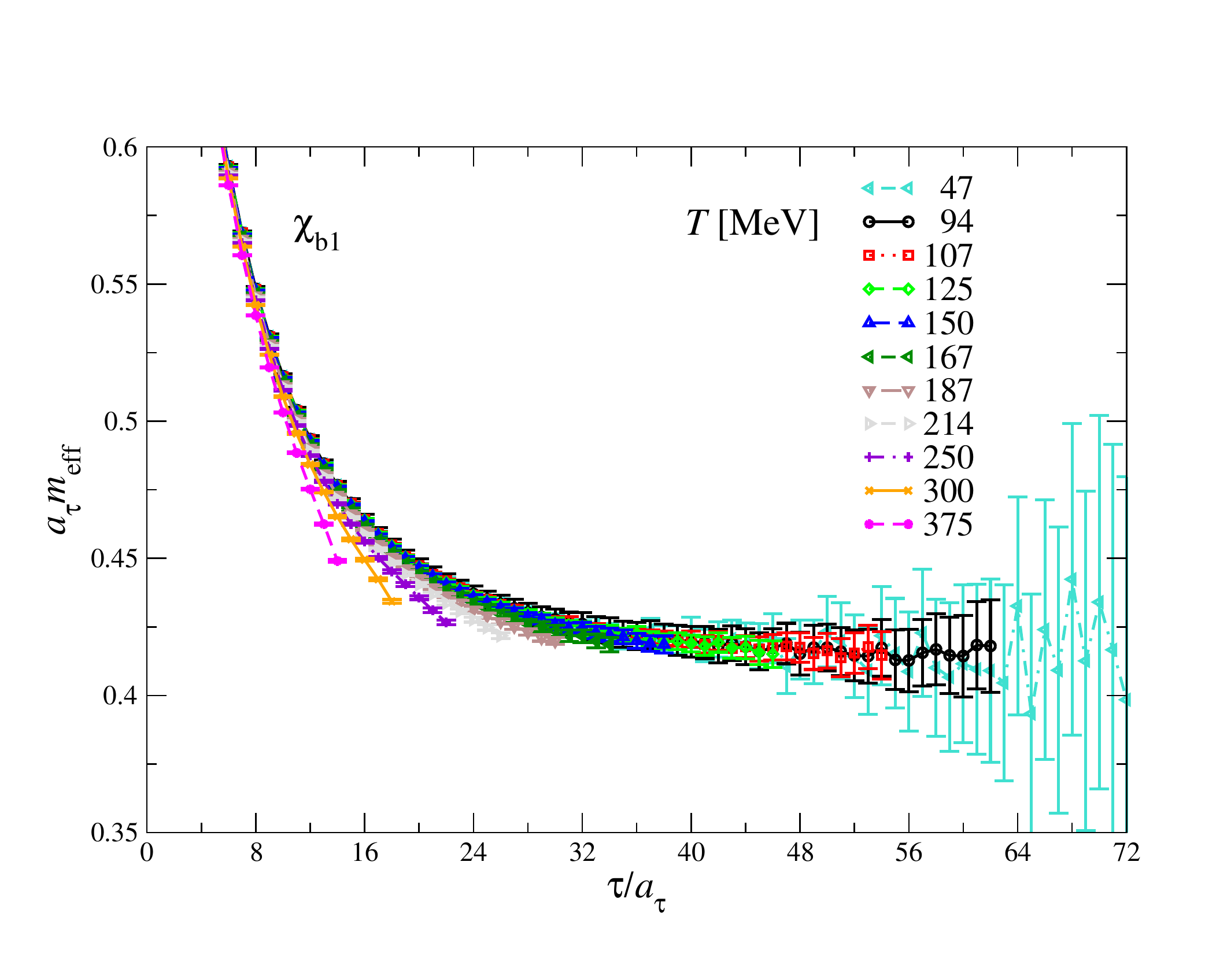}
	\caption{Effective mass in the $\Upsilon$ (left) and $\chi_{b1}$ (right) channels, for a range of temperatures. The resulting masses at the lowest temperature are $m(1S) = 9.460$ GeV and $m(1P) = 10.005$ GeV respectively. The latter is somewhat on the heavy side, which requires further investigation.
		\label{fig:mass}}
\end{center}
\end{figure}

We follow the approach discussed in detail in Refs.\ \cite{Aarts:2011sm,Aarts:2014cda}. Bottomonium correlators are generated by solving the NRQCD evolution equations for the $b$ quark, whose mass is tuned via the non-relativistic dispersion relation of the spin-averaged 1S groundstate. Hadronic spectral quantities are determined up to an overall energy constant $E_0$, which is fixed via the experimental value of the $\Upsilon(1S)$ mass. We find that $E_0=7464.5$ MeV. 
Some results in the  $\Upsilon$  ($S$ wave) and $\chi_{b1}$  ($P$ wave) channels are shown here. In Fig.\ \ref{fig:ratio}  we present the ratio of Euclidean correlators at a given temperature $T$ to the correlator at $T_0=47$ MeV, the lowest temperature on which the bottomonium correlators were computed, allowing for a first indication of thermal effects. Fig.\ \ref{fig:mass} contains  temperature-dependent effective masses. 
In agreement with earlier observations \cite{Aarts:2011sm,Aarts:2012ka,Aarts:2013kaa,Aarts:2014cda,Kim:2014iga}, we note significant temperature dependence in $P$ waves, not present in $S$ waves. 
 In fact, there is quantitative agreement with the Gen 2 results \cite{Aarts:2014cda}, indicating only a weak pion mass dependence.

\section{Kernel Ridge Regression}

In order to gain a more detailed understanding of spectral modification in a thermal medium, it is of interest to extract the spectral function $\rho(\om)$, related to the Euclidean correlator $G(\tau)$  via\footnote{The form of the kernel $K(\tau,\om)$ depends on the problem; for NRQCD it is the simple exponential given here \cite{Aarts:2010ek,Aarts:2011sm}.}

\be
	G(\tau) =  \int_{\omega_{\rm min}}^{\omega_{\rm max}} \frac{d\om}{2\pi}\, K(\tau, \omega) \rho(\omega),
	\qquad\qquad\qquad
	 K(\tau, \omega) = e^{-\om\tau}.
	 \label{eq:grho}
\ee
As is well known, determining $\rho(\om)$ from a numerically computed correlator $G(\tau)$ at a finite number of temporal lattice points  is an ill-posed inversion problem. This has been addressed using  the Maximum Entropy Method (MEM) and other approaches (see Ref.\ \cite{Rothkopf:2019ipj} and references therein), but these methods are not yet entirely satisfactory as they suffer from
systematic uncertainties.

In the last decade the field of machine learning has undergone significant development, with considerable success in image classification and reconstruction, given incomplete and noisy input data. Hence it makes sense to investigate whether machine learning methods can be applied to spectral reconstruction in the context of lattice QCD.
If successful, this alternative approach can then be used to support the results of other methods such as MEM, with the potential to surpass it in the future. While there are many methods available, here we will focus on Kernel Ridge Regression (KRR), which has been used in Ref.\ \cite{krr} to address the analytic continuation problem in quantum many-body physics, using mock data only. An alternative Deep Neural Network approach has been tested on mock data in Ref.\ \cite{Kades:2019wtd}. In this contribution, we will apply KRR to actual lattice data. 

Kernel Ridge Regression, like linear regression, is used to determine a mapping between two datasets. However, KRR uses sets of functions as opposed to sets of values as in linear regression. In our context the KRR model is used to map a set of correlators to a set of spectral functions, i.e.\ to establish the relation $(G, \rho)$.
The method for training a KRR model can be separated into three steps: generation of training data, determination of a parameter matrix $\mathbf{\alpha}$ which relates the input and output data, and lastly selecting the optimal KRR hyperparameters. The final step is optional, but is done to improve predictions. We note here that the generation of training data is straightforward: one can construct an essentially unlimited set of spectral functions by combining multiple (narrow and broad) peaks and a continuum contribution at large energies, while imposing an upper limit $\om<\om_{\rm max}$ arising from the finite lattice cutoff \cite{Aarts:2011sm}. 
In practice we generated the training set by combining Gaussians of various widths, while enforcing the constraints $\rho(\om=\om_{\rm min}) = 0$,  $\rho(\om) > 0$,  and 
\be
 \int_{\omega_{\rm min}}^{\omega_{\rm max}} \frac{d\om}{2\pi}\, \rho(\omega) = G(\tau=0) = \int d^3x\,\, S(\vecx),
 \label{eq:S}
 \ee
 where $S(\xv)$ is the source used in the NRQCD formulation, see Ref.\ \cite{Aarts:2011sm} for details.
Given these spectral functions, the associated Euclidean correlators follow from the easy-to-evaluate integral (\ref{eq:grho}),
which is performed either by integration or by summation, after discretising the $\om$ interval using $N_\om$ points.

One issue with spectral reconstruction, i.e.\ the inversion of Eq.\ (\ref{eq:grho}), is the mismatch in the number of points $N_\om$ needed to describe the spectral function $\rho(\om)$, say $\mathcal{O}(1000)$, and the number of points available from the correlator $G(\tau)$, which is $\mathcal{O}(N_{\tau})$. This can partly be avoided by expressing $\rho(\om)$ in terms of an incomplete set of basis functions, for instance by employing the decomposition familiar from MEM \cite{Aarts:2007wj}, 
\be
	\rho(\omega) = m(\om) \exp \sum^{N_a}_{k=1} a_k f_k(\omega).
\label{eq:rhoa}
\ee
Here $m(\om)$ is a polynomial in $\om$, $a_k$ ($k=1,\ldots N_a$) are a set of coefficients, and $f_k(\omega)$ form the set of basis functions, satisfying
\be
\int_{\omega_{\rm min}}^{\omega_{\rm max}}  d\om\, f_k(\om)f_l(\om)= \delta_{kl}.
\ee
 For the results shown here, we use a constant $m(\om)= m_0$. The choice of basis functions is inspired by the construction used in MEM \cite{Aarts:2007wj}. For this we consider the kernel $K(\tau_k,\om_n)$, after discretising both $\tau$ and $\om$, as an $N_a\times N_\om$ matrix, and generate the basis functions by taking the SVD of $K(\tau_k, \omega_n)$. The crucial difference with MEM is that $N_a$ does not depend on the number of time slices available at a given temperature, $N_{\tau}$, but is a free parameter chosen such that $N_\tau  \ll N_a \ll N_\om$. Changing $N_a$ allows for checks of stability. Given a spectral function in the training set and a constant $m(\om)=m_0$, the coefficients $a_k$ are then determined by
 \be
a_k =  \int_{\omega_{\rm min}}^{\omega_{\rm max}} d\om\, f_k(\om)\log\left[ \frac{\rho(\omega)}{m_0}\right].
\ee
KRR will now establish the map between $G(\tau)$ and the set $\{a_k\}$.

To train the model, we note that KRR is a combination of two techniques. The first of these, the kernel method, can be treated as a generalised form of linear regression. In the linear case $y = w^T\phi(x)$, where $y$ are the target data, $\phi(x)$ is a vector of functions of the input data $x$, and $w$ is a vector of parameters \cite{bishop}. The cost function to be minimized is  $E=\half(y - w^T\phi(x))^2$.
In our application, the  input data are a training set of Euclidean correlators $G_i(\tau)$ ($i=1,\ldots, N_{\rm train}$). Rather than using these directly, they are combined in a $N_{\rm train}\times N_{\rm train}$ matrix $\mathbf{C}$, with 
\be
C_{ij} = \exp \left( -\frac{1}{\sigma^2} \sum_n \left[G_i(\tau_n) - G_j(\tau_n)\right]^2\right),
\ee
where the sum goes over all time slices.\footnote{The actual form of the kernel function can be varied with the only restriction being that it determines a difference between the input data \cite{bishop}. It should not be confused with the integral kernel appearing in Eq.\ (\ref{eq:grho}).}
 The quantity $\sigma$ is a hyperparameter, which sets a correlation length in the space of correlators.
The target data, i.e.\ the spectral functions, are encoded in a  $N_{\rm train}\times N_a$  matrix $\mathbf{Y}$, in which each row contains the $N_a$ coefficients determining $\rho_i(\om)$ according to Eq.\ (\ref{eq:rhoa}). Finally, the input and target data are assumed to be related according to 
\be
\mathbf{Y} = \mathbf{C}\mathbf{\alpha},
\ee
which defines the $N_{\rm train}\times N_a$ matrix $\mathbf{\alpha}$ as the equivalent of $w$ in linear regression. The aim of the training stage is to determine this $\mathbf{\alpha}$.

The second ingredient in KRR is ridge regression, which adds an additional term in the cost function, proportional to the square of the parameters, to prevent overfitting. This gives the following cost function
\be
  E(\mathbf{Y, C, \alpha} ) = \half(\mathbf{Y} - \mathbf{C}\mathbf{\alpha})^2 + \half\lambda\mathbf{\alpha}^T\mathbf{C}\mathbf{\alpha},
\ee
where $\lambda$ is the second hyperparameter, used to regularise the influence of the additional term. Minimising this cost function with respect to $\mathbf{\alpha}$ then determines the optimal parameter matrix, for given $\sigma$ and  $\lambda$, as
 \be
\mathbf{\alpha}_{\rm opt} = \left( \mathbf{C} + \lambda \mathbf{I} \right)^{-1} \mathbf{Y}.
\ee
The hyperparameters are determined via a cross-validation procedure. After this training stage, it is possible to make predictions for a spectral function ($\mathbf{Y'}$) given an actual Euclidean correlator using
\be
\mathbf{Y'} = \mathbf{C'}\mathbf{\alpha}_{\rm opt},
\ee  
where $\mathbf{C'}$ is determined from the squared difference between the NRQCD correlator and the training correlators, i.e.\ it is a matrix of size $1\times N_{\rm train}$.

\begin{figure}[t]
\begin{center}
	\includegraphics[width=0.49\linewidth]{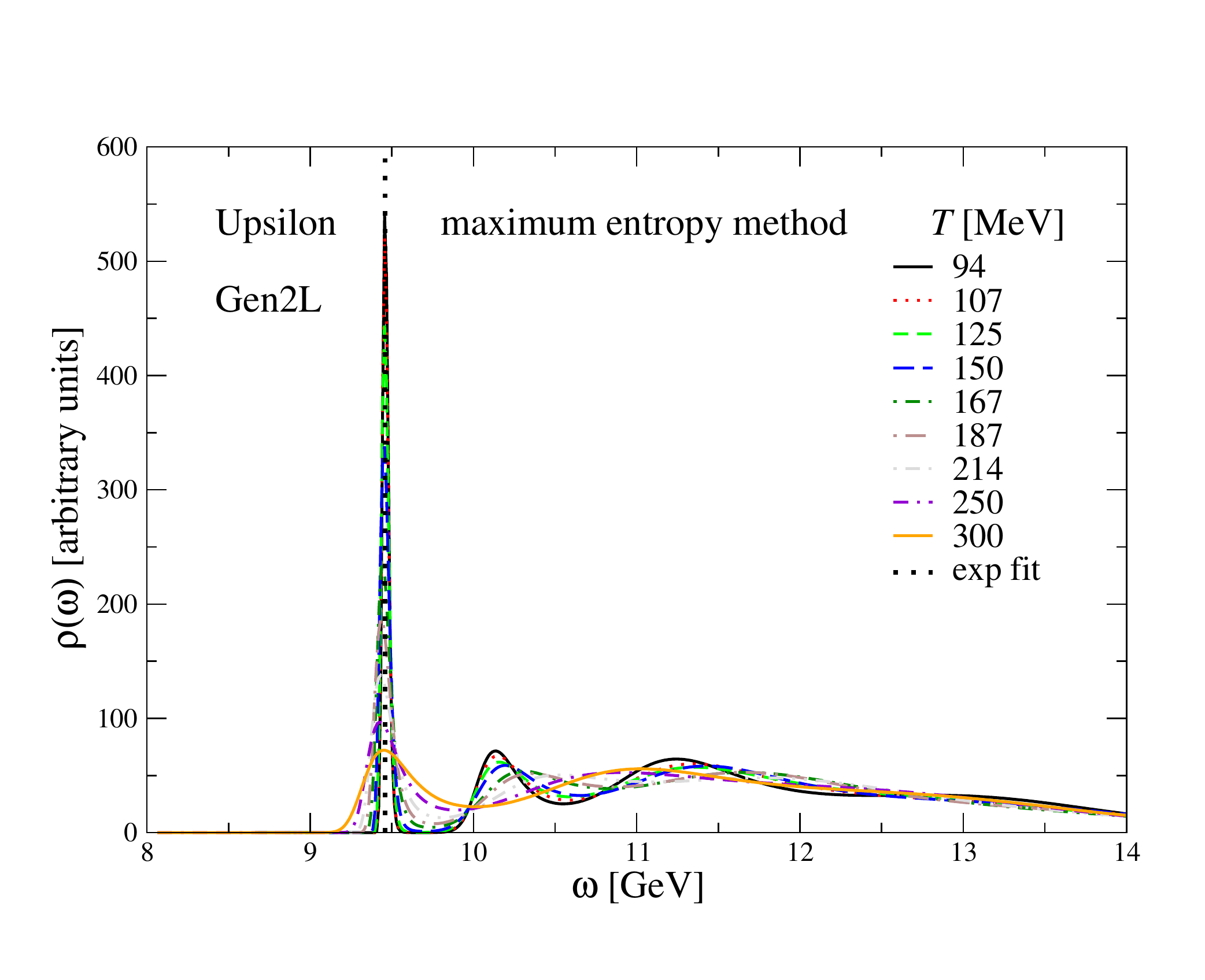}
	\includegraphics[width=0.49\linewidth]{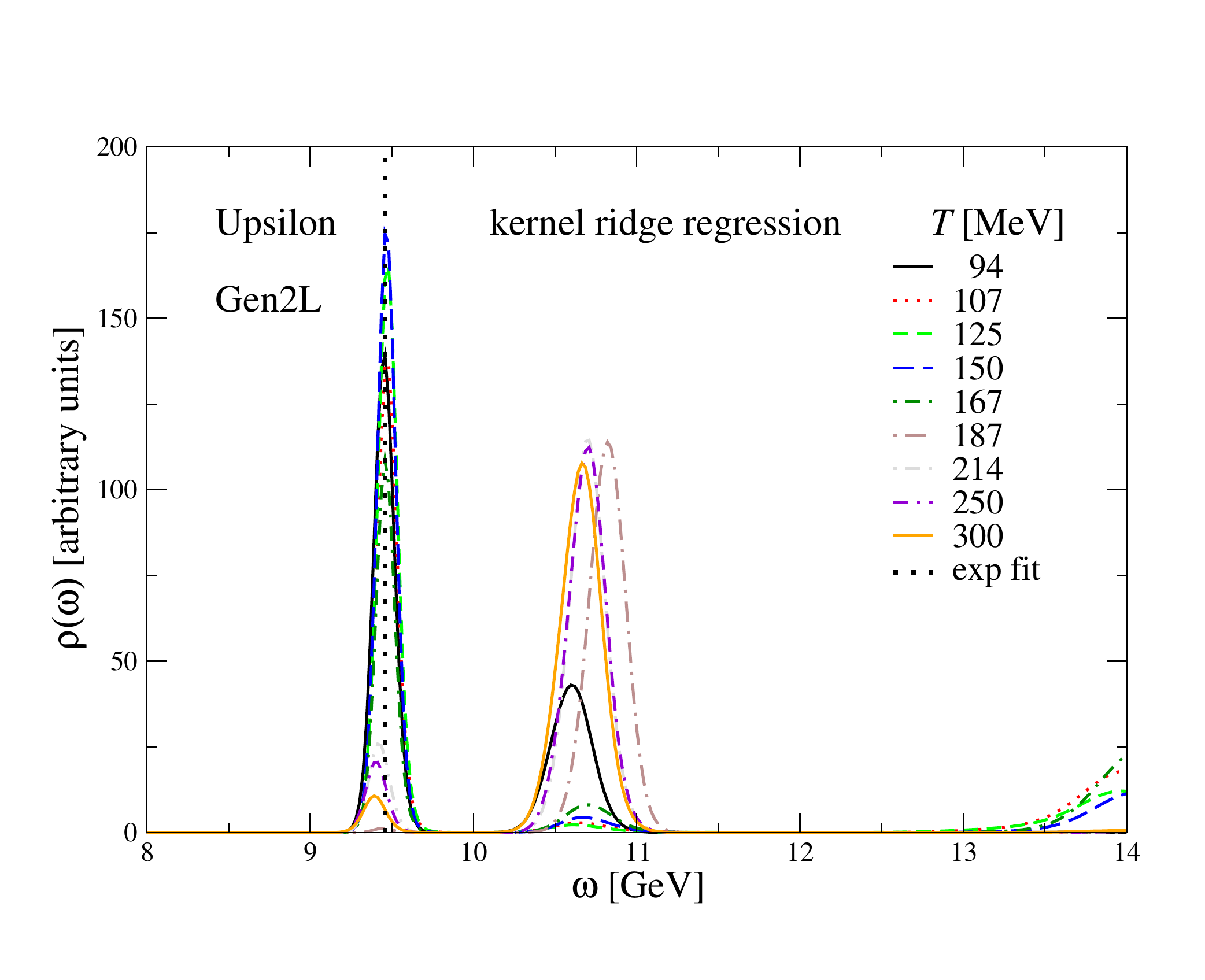}
	\caption{Spectral functions, obtained with the Maximum Entropy Method (MEM, left) and Kernel Ridge Regression (KRR, right) using $N_a=80$ basis functions, in the $\Upsilon$ channel for a number of temperatures, in Gen 2L. The dotted line indicates the mass at the lowest temperature, obtained used a conventional fit.
	\label{fig:KRR}}
\end{center}
\end{figure}

First preliminary results of this approach are presented in Fig.\ \ref{fig:KRR}, in the $\Upsilon$ channel, comparing MEM (left) and KRR (right). 
The latter are obtained using $N_a=80$ coefficients to parametrise the spectral functions, at all temperatures. The size of the training set is $N_{\rm train} = 9000$. Training is repeated at every temperature, using the same training set $\{\rho_i(\om)\}$, but for correlators appropriate at that temperature, i.e.\ with given $N_\tau$. We note a clear first peak, whose position coincides with the mass obtained using a standard exponential fit at the lowest temperature, indicated with the dotted line. As the temperature increases, the size of the first peak drops above $T=187$ MeV, in both MEM and KRR. Given the difference between the two approaches, this is an interesting observation. 
The area under the spectral function is the same at all temperature, see Eq.\ (\ref{eq:S}); hence a reduction of the size of the first peak leads to  more spectral weight at higher energies, both in MEM and KRR. In the case of KRR this leads to a more pronounced second peak. However, the second peak is at this moment not yet reliable and requires further study.

\section{Summary}

We have shown new results for bottomonium on the FASTSUM Gen 2L ensembles, finding quantitative agreement with previous Gen 2 results. Further analysis will be carried out shortly, including the application of 
the methods developed in Ref.\ \cite{Larsen:2019bwy}.
We have presented  preliminary results for spectral reconstruction using  a machine learning technique, namely Kernel Ridge Regression, applied to actual lattice QCD data for the first time.
The position of the first peak was seen to coincide with the one obtained from conventional fitting and from MEM, with a drop of the peak size at higher temperature seen both with KRR and MEM.  Even though much remains to be done, we take these first results as a positive stimulus for the future.

 \vspace*{0.5cm}
 
\noindent
{\bf Acknowledgments.}
  We are grateful for support from STFC via grants ST/L000369/1 and ST/P00\-055X/1, the Swansea Academy for Advanced Computing, SNF, ICHEC, COST Action CA15213 THOR, and the European Research Council (ERC) under the European Union's Horizon 2020 research and innovation programme under grant agreement No 813942.
Computing resources were made available by HPC Wales, Supercomputing Wales, and PRACE via access to the Marconi-KNL system hosted by CINECA, Italy.
This work used the DiRAC Extreme Scaling service and the DiRAC Blue Gene Q Shared Petaflop system at the University of Edinburgh, operated by the Edinburgh Parallel Computing Centre on behalf of the STFC DiRAC HPC Facility. This equipment was funded by by BIS National E-infrastructure capital grant ST/K000411/1, STFC capital grant ST/H008845/1, and STFC DiRAC Operations grants ST/K005804/1 and ST/K005790/1, and also through BEIS capital funding via STFC capital grant ST/R00238X/1 and STFC DiRAC Operations grant ST/R001006/1. 
DiRAC is part of the National e-Infrastructure.

\end{document}